\documentclass[a4paper]{ESASPStyle}

\def\approxlt{\raise3pt\hbox{${\scriptstyle<}$}
           \kern-6pt\lower1.1pt\hbox{${\scriptstyle\sim}$}}
\def\approxgt{\raise3pt\hbox{${\scriptstyle>}$}
           \kern-6pt\lower1.1pt\hbox{${\scriptstyle\sim}$}}

\usepackage{epsfig}
\newlength{\lenA} %
\newcommand{\mnras} {MNRAS}
\newcommand{\aap} {A\&A}

\begin{document}

\title{Inferring information about rotation from stellar oscillations }

\author{M.J. Goupil\inst{1} \and R. Samadi\inst{1,2} \and J. Lochard\inst{1} 
\and W.\,A.  Dziembowski\inst{3,4}\and A. Pamyatnykh\inst{4,5}}
 \institute{Observatoire de Paris- LESIA, 92190 Meudon, France 
  \and Departamento de Matem\'atica, FCTUC, Coimbra, Portugal
  \and Warsaw University Observatory, Al. Ujazdowskie 4, 00-478 Warsaw, Poland
  \and Copernicus Astronomical center, Bartycka 18, 00-716, Poland
  \and Institute of Astronomy, Pyatnitskaya 48, 109017 Moscow, Russia
}

\maketitle 

\begin{abstract}

 The  first part of this paper aims at  
illustrating the intense scientific activity in the field of stellar rotation 
although, for sake of shortness, we  cannot be exhaustive nor 
give any details. The 
second part is devoted to the rotation as 
a pertubation effect upon oscillation frequencies. The discussion focuses on   one specific 
example: the $p$-modes frequency small separation  which provides information about properties 
of the stellar inner layers. It is shown that 
the   small separation can be affected by
rotation at the level of 0.1-0.2 $\mu$Hz for a 1.4 $M_\odot$ model rotating 
with an equatorial    velocity of 20 km/s at the surface. This is of the same order of magnitude as the expected precision 
 on frequencies with a 3 months observation
and must therefore be taken into account. We show however that it is possible 
to recover the small separation free of  these contaminating effects of rotation, 
provided enough high quality data are available as will be with space seismic missions such as Eddington.

\end{abstract}

\section{Introduction}
\label{sec:intro}
  
\begin{figure*}
  \begin{center}
\resizebox{14cm}{!}{   \rotatebox{-90}{  \epsfig{file=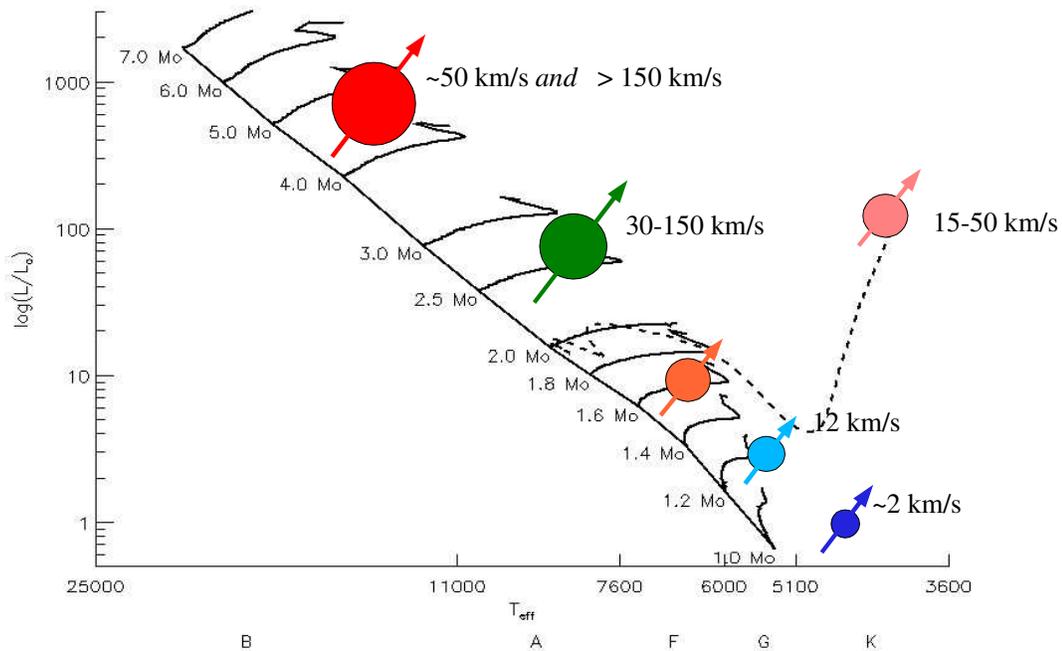} } }
  \end{center}
\caption{HR diagram indicating the range of rotational velocity for  stars  with various masses and ages. \label{fig1}}
\end{figure*}

The importance of the role of rotation in stellar evolution  has long been recognized.
 Effects of  rotation is nowadays taken into consideration in many stellar studies
and is  often  included in the modelling of stars, their structure and evolution.
Stars of different masses and ages have rotational velocities which span  a wide range (Figure~\ref{fig1}). 
Determining the dependence of the rotational properties of stars with mass and age  is an active field.
 One important issue is to understand and get 
 a global picture of the evolution of the rotation rate inside a star:

$\bullet$ with age from the PMS stage (Matthieu, 2003) up to the  stage of compact objets (Woosley, Heger, 2003)

$\bullet$ with mass from early O-A-B type stars down to late F-G-K type stars

The global picture is essential for understanding observed properties of open clusters or, 
to give another example, for explaining the origin of  of gamma ray bursts which  progenitors must be massive rotating stars. 
Then  the goal is to  explore the 
consequences of the evolution of the angular momentum  on the evolution of stars and galaxies. 
Progress in understanding evolution of angular momentum will result in improving the age determinations
 based on isochrone fitting,  to understand the  evolution of massive stars, 
the chemical evolution of galaxies. As a  feedback,  this will provide  
improved theories of   transport and evolution of stellar angular momentum in stellar conditions.

Efforts to measure stellar rotation rates  
- at least projected rotational velocities $v \sin i$-  
for many stars of different types   have recently been intensified.

\begin{figure}[ht]
  \begin{center}
    \epsfig{file=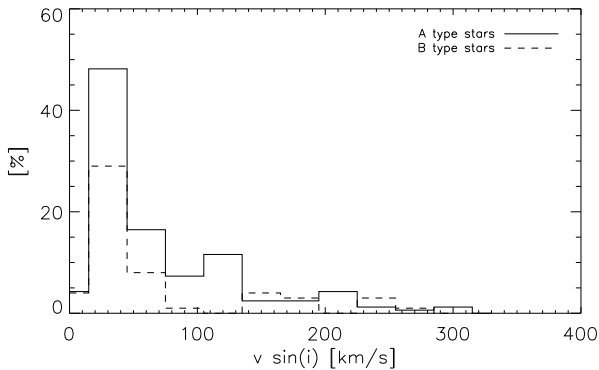, width=\lenA}
  \end{center}
\caption{Histograms  showing the relative  number of stars 
found with a $v \sin i$   in a given range.    
Data are from Royer et al.2000.  
The solid line represents A stars (164 objects), 
the dashed line corresponds to B stars (53 objetcs)  \label{fig2}}
\end{figure}

A homogeneous sample of $v \sin i $ measurements has been obtained for A-B type stars (Royer et al. 2002a,b). 
The resulting  histograms in 
Figure~\ref{fig2} show that the bulk of stars is found in the range 20 -100 km/s  
indicating that   fast rotation is common among A-B stars; $v \sin i$ values of   200-300 km/s are not infrequent.

It is necessary  to determine the projected 
rotational velocity of $\delta$ Scuti stars before any seismological study of these rapidly rotating  variable stars  
be possible.  This is  particularly important for 
$\delta$ Scuti stars in  open clusters as for cluster  stars,   
other parameters, chemical composition, age -  are known. 
Projected rotational velocities for a few $\delta$ Scuti stars  have been measured in   
 Praesepe or NGC 6134  cluster (Rasmussen et al., 2002) for instance.

\begin{figure}[ht]
  \begin{center}
    \epsfig{file=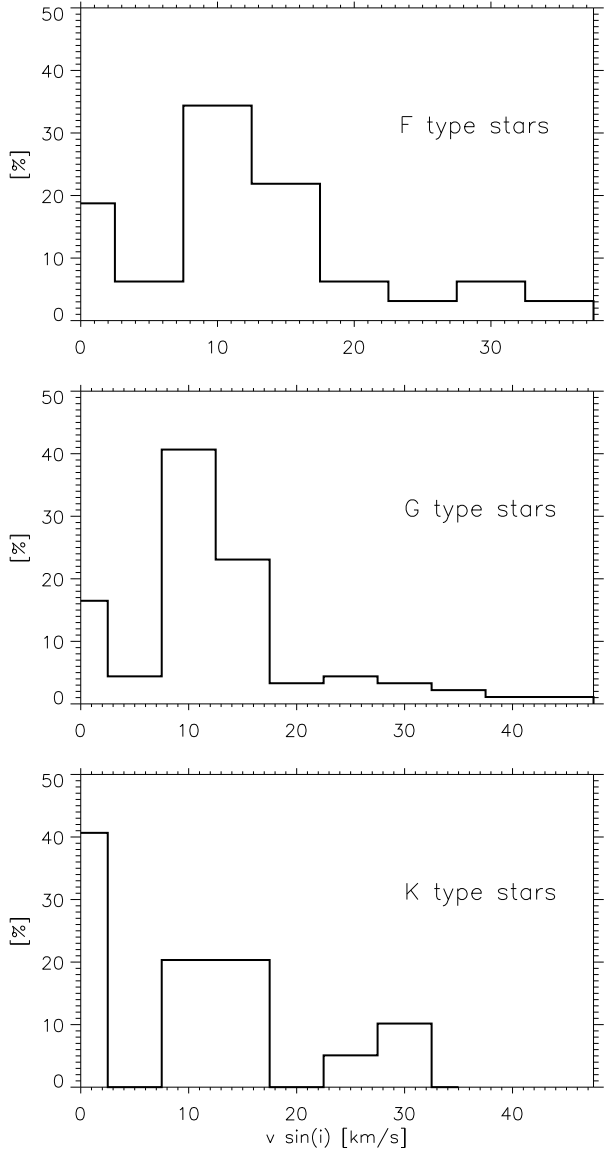, width=\lenA}
  \end{center}
\caption{Histograms as in Fig.2 but for F type stars (32 objects), 
G type stars (91 objects), K type stars (19 objects).
     Data are from Cutispoto et al. 2002. \label{fig3}}
\end{figure}

In order to study connections between rotation, X ray-emission, 
lithium abundance and stellar activity (Cutisposto et al (2003)), Cutisposto et al (2002)  have determined
 the $ v \sin i$ of  a large sample of  F-G-K type objects,  in the solar neighborghoud, selected for 
their high $v \sin i$ (Figure~\ref{fig3}). 
With the purpose of determining possible latitudinal differential rotation and its correlation 
with stellar activity  in stars other than the Sun, Reiners and Schmidt (2003) have obtained 
projected rotational velocities for a sample of field 
F-G-K type PMS and MS stars.  Of interest here is that, 
although these are typically  much slower rotators than A-B type stars, 
 $v \sin i $ values exceeding  15 km/s  are not uncommon,  
as we may see  in Figure~\ref{fig3}.

Determination of  projected rotational velocities 
in very young cluster stars  is also carried out (Herbst et al., 2001; Terndrup et al., 2002) 
as well as measurements of  rotation periods and latitudinal differential rotation 
(Petit et al, 2002; Cameron et al. 2002; Barnes, 2003) 
with the aim of providing constraints in the modelling of the evolution of angular momentum 
during the very early phases of star formation and PMS stages.

\section {Rotation in stellar models}

\subsection{ Stellar evolutionary models with rotation}
Inclusion of rotation in calculations of 
1D evolutionary models  is now performed by several groups.

Meynet, Maeder (1997) (see also Meynet, Maeder 2000 and ref. therein) 
 have developed an evolutionary  code which is  based on a modified  version 
of Zahn's prescription 
 which treats rotation effects as  diffusion and  advection processes (Maeder, Zahn, 1998).
 The modelling works have stimulated new  theoretical studies about the rotation rate profile in the
 $\mu$-gradient zone  (Talon, Charbonnel 2003; Palacios et al, 2003) 
and  transport of angular momentum by  gravity waves (Talon et al, 2002).
The observational constraints are the abundances of chemical elements 
at the surface of stars which can give hints  about the rotationally internal mixing processes. 
A number of calculations with the Geneva 'rotating' code   
has  enabled   investigations of several problems such as 
He overabundances in O stars (Maeder, Meynet, 2000a), more generally  evolution of -
 and yields by - rotating 
 massive   stars (Maeder, Meynet, 2003; Meynet et al., 2003).

The Yale group  studies rotating  solar type stars with models built following 
 the Endal and Sofia (1976) approach   
which includes a somewhat different  prescription 
for the  evolution of the angular momentum and rotationally induced mixing, 
(Sills et al. 2000). For instance,  Barnes et al. (2001) 
investigate the possibility that disk locking can 
efficiently spin down young stars  
as observed in young clusters. Barnes (2003) compiles available 
 observations of rotation 
periods of  cluster stars, classifies the stars as belonging to different sequences and  
 proposes an interpretation of these observations 
as related to the properties of stellar magnetic fields.   

Using  Endal and Sofia's code,  another group also studies the influence of rotation on the evolution of 
massive stars (Heger et al. 2000a,b). Observations seem to confirm the model prediction 
which states that  main sequence OB stars  are
already mixed  which causes boron to be depleted without nitrogen being affected (Venn et al. 2002).

All these models show that the evolutionary tracks are significantly 
modified by fast rotation 
and any conclusions drawn from the position of a star in the HR diagram  are then affected,
particularly ages, isochrones  and masses, amount of overshooting 
(Maeder, Meynet 2000b;  Heger, Langer, 2000a; 
Maeder, Meynet, 2003;   Palacios et al, 2003)

Calculation of binary evolution including the  rotation of both components, 
transport of angular momentum and tidal processes 
are also now being  carried out. Rotation appears to be able to decrease significantly the efficiency of the  
accretion process in some cases; it   has a strong effect on the evolution of  massive binary systems, 
playing an important role in the 
the formation of a black hole in a binary system for instance 
or in the evolution of progenitors of type I supernovae
(Langer et al,  2003 and ref. therein)

\subsection{ 2D-3D modelling}

In order to provide a prescription in 1D models as close to the reality as possible, 
one needs a better understanding

\noindent of  transport and mixing processes in the interior of stars; 
this can come from 2D and 3D calculations.

Fully 2D rotating models of  stars show for instance that the shape of a 
convective core changes with increasing rotation rates from a spherical to an 
aligned with the rotation axis geometry  (Deupree, 2001).

3D modelling of a rotating spherical shell which is a simplified representation of the 
convective core of a rotating A type star 
shows that indeed  rotation strongly influences the convective motions; it
provides clues about the properties of overshooting  which appears to 
vary with latitude creating  asymmetries in the extent of the penetration (Brun et al., 2003).

3D MHD   computations of a rotating spherical shell representing the solar 
outer convective envelope  are performed in order to study 
the relations between differential rotation, turbulent convection  and  stellar magnetic activity 
with the conclusion that rotation and convection 
can generate magnetic activity but for these to be cycles, 
the existence of a tachocline might be crucial (Brun, 2003).

\section {Stellar rotation and oscillations}

Stellar oscillation frequencies $\nu_{n,\ell,m}$ are labelled with 3 characteristic numbers: 
the radial order  $n$, the degree $\ell$ and the azimuthal order $m \in [-\ell,\ell]$. 
In absence of rotation, the frequencies are $2\l+1$ degenerate   $\nu_{n,\ell,m}= \nu_{n,\ell,0}$  for 
  $m \in [-\ell,\ell]$. Rotation breaks the azimuthal symmetry and lifts the $m$-degeneracy.

If rotation is slow enough and its angular rate (hereafter denoted $\Omega$) is independent of latitude, then
 the first order perturbation method in $\Omega$ predicts an equidistant frequency splitting, $\delta_{n,\ell}$,
between consecutive $m$-components within each $(n,\ell)$ multiplet. For uniform rotation, the values of 
  $\delta_{n,\ell}$ are proportional to $\Omega$ and allow to measure the rotation rate, free of the $\sin i$ uncertainty.
In the case of a shellular rotation, $\Omega=\Omega(r)$, the $\delta_{n,\ell}$ 
provide mode-dependent mean values of the interior rotation rate. If we define the symmetric frequency splitting as 
\begin{equation}
\delta_{nl} = {\nu_{n,\ell,m}-\nu_{n,\ell,-m} \over 2 m}
  \label{eq:equation1}
\end{equation}
then the connection between such splitting and the interior rotation remains accurate up to $\Omega^2$ because the terms 
$\propto \Omega^2$ are even in $m$.

\subsection{Seismic measurements of rotation rates }

Seismic measurements of rotation deduced from frequency splittings, i.e. from $\delta_{nl}$ (Eq.1), 
  are now available for a few individual stars.
For instance,  seismic measurements of the rotation rates  
of variable white dwarfs  seem to be the easiest approach
  and possibility of determining 
a nonuniform rotation for these stars  has been explored  (Kawaler et al., 1999).

For other types of pulsating stars, 
   the rotation rate deduced from seismic measurements is model dependent 
as it usually relies on a mode identification which is itself  
model dependent. Today  $\beta$ Cephei stars are the best
 candidates to provide reliable mode 
identifications,hence the deduced rotation rates,   
which we can trust (Thoul et al. 2003; Handler et al., 2003).
For $\delta$ Scuti stars, the process of mode identification remains difficult and this keeps one from
determining   their rotation rates with
 enough confidence, except perhaps for FG Vir for which line-profile and light 
variations are available (Mantegazza, Poretti 2002).

For solar-like  oscillating stars, it is also proposed to measure directly the total 
angular momentum (Pijpers, 2003); this requires 
a large set of high quality frequencies as will be obtained by the forthcoming 
seismological space missions (MOST, COROT, Eddington)

\subsection{Rotational effects on mode excitation }
Rotation affects  driving and damping of  oscillation modes
to different extent depending on the type of stars and of excited modes
 (Lee 1998; Ushomirsky, Bildsten, 1998; Saio et al.,  2000; Lee, 2001). 
Correlations appear to exist between the amplitudes of  the excited modes and 
the $v \sin i$ of $\delta$ Scuti stars in Praesepe cluster (Suarez et al., 2002); 
this is an important issue 
regarding possible selection effects which are responsible for the complicated 
patterns of oscillation frequency spectra of these stars and  further work is necessary to confirm it. 
The  Eddington space mission  is well suited for such statistical  studies 
from which we expect more complete oscillation spectra and the value  of the rotational velocity 
$v$ instead of the projected one $v \sin i$ for a large sample of stars. 

\subsection{Rotational effects on frequencies}
 
One  must however be aware that rotation can also hinder  some seismic studies (Dziembowski \& Goode, 1992 (DG92)).
  
For  stars oscillating with opacity driven modes such as $\delta$ Scuti and $\beta$ Cephei variables 
which oscillate  in the low frequency- low radial order range- and as these stars rotate fast, 
third order effects in $\Omega$ as measured by  $\Omega^3/\nu^2_{n,\ell,0} $ range in the interval 
$\sim 0.01- 0.5$ $\mu$Hz 
and must therefore generally  be taken into account   in calculating the splittings
(Goupil et al., 2001; Dziembowski et al., 1998; Goupil \& Talon, 2002;  Pamyatnykh,  2003)

For very fast rotating stars as for instance some  $\delta$ Scuti stars known to have a 
$ v \sin i\sim  250$ km/s, even  a third order perturbation is no longer valid and a
nonperturbative approach is the only appropriate one. Such investigations have been 
initiated  with a special attention to low frequency gravity modes 
(Clement 1981-1998; Lee \& Saio, 1987; Dintrans \& Rieutord, 2000) which can apply to $\gamma$ Dor stars.
In order to consider acoustic modes, the effect of the centrifugal force on the equilibrium state 
of the star and on the wave motions must also be included. Based on the numerical approach of 
Rieutord \& Valdettaro (1997) which allows to include a large number of spherical harmonics in the description 
of each individual oscillation mode without being  numerically too time consuming,
 Ligni\`eres et al. (2002) and Ligni\`eres (2003, in prep)  
have calculated the oscillation frequencies of a rotating polytrope with an oblateness up to  $\epsilon=0.15$. 
Results show that the mode amplitude inside the star even at this relatively small oblateness 
may become a quite complicated function of $r$ and $\theta$, significantly departing from a 
 single spherical harmonic description. These calculations also confirm that 
the small separation is affected  particularly when it involves  mixed modes  (see also Sect.3.4 below). 

\vskip 0.3 truecm
For solar-like stars, 
the small separation between  $\ell=0$ and a $\ell=2$ modes as well as $\ell=1$ and $\ell=3$ 
can become  so small that rotation, even at a modest rate of  15  km/s, leads to a significant 
departure from the  single spherical harmonics representation of individual modes.

In the framework of a perturbative approach, this means that the 
radial mode $\ell=0$  is no longer purely radial but is contaminated with a $\ell=2$ contribution and
 conversely the $\ell =2$ mode is contaminated  with a $\ell=0$ contribution. 
This has severe consequences already stressed in  Soufi et al. (1998),  Dziembowski (1997), 
Dziembowski\& Goupil (1998)

\subsection{Rotation as a perturbation  }

The small separation $d^{(0)}_{n,\ell}$ is defined as:
\begin{equation}
d^{(0)}_{n,\ell} =  \nu^{(0)}_{n,\ell}- \nu^{(0)}_{n-1,\ell+2} 
\end{equation}
where  $ \nu^{(0)}_{n,\ell}$ is the eigenfrequency of mode with degree $\ell$ and radial order $n$ 
in absence of rotation.  
The small separation $d^{(0)}_{n,\ell} $ 
is known to  provide information about the structural properties of the inner 
layers of the star (Gough 1998; Roxburgh, Vorontsov, 1998 ).
However, rotation modifies  the structure of the star  and its oscillation frequencies. 
When  rotation is fast enough,
 the frequencies can be modified to an extent that the 
small separation $d^{(0)}_{n,\ell}$ is significantly affected.
 
This is illustrated below.
Only the symmetrical part of the centrifugal distorsion 
is directly  included in the stellar model as an effective gravity, 
no mixing due to rotation is taken into account. A uniform rotation is assumed 
 and evolution of the stellar model is performed assuming conservation of total angular momentum. 
Other effects of rotation  on the oscillation frequencies are 
taken into account by means of a perturbation method. 
For moderate rotation rates as are expected for most solar-like oscillators,
 cubic order terms, $O(\Omega^3/\nu^2)$,  in the perturbation ($\sim 10^{-6}-10^{-5} \mu$Hz) 
can be neglected and only second order effects in $\Omega$ are considered.
In that case, one obtains,  for each $m$ component of a given multiplet of modes $\ell,n$:
\begin{equation} 
\nu_{n,\ell,m} = \nu_{0,n,\ell}+m~{\cal C}_{n,\ell}+ {\cal D}_{n,\ell,m}
\end{equation}
where
$\nu_{0,n,\ell}$ is the eigenfrequency of ($n,\ell$) mode of a 
stellar model built with rotation through an effective gravity;  
 ${\cal C}_{n,\ell} $ represents the first order (Coriolis) correction due to rotation and reduces to 
$m \Omega (C_{n,\ell}-1) $ for a uniform rotation in the observer frame 
with $C_{n,\ell}$ the Ledoux constant (DG92). 
The quantity ${\cal D}_{n,\ell,m}$ is a $O(\Omega^2)$ contribution which takes into account
the effect of the nonspherically symmetric  centrifugal distorsion of the stellar model 
and the second order correction due to Coriolis force.

For nondegenerate  modes in a rotating star, the small separation is then given by
\begin{equation} 
d_{n,\ell,m} \equiv \nu_{n,\ell,m}- \nu_{n-1,\ell+2,m}
\end{equation}

From Eq.(3), the small separation of non degenerate $m=0$ modes can be written as:
\begin{equation} 
d_{n,\ell,0}
 = \Delta_{n,\ell,0} +  \Bigl({\cal D}_{n,\ell,0}-{\cal D}_{n-1,\ell+2,0}\Bigr)
\end{equation}
where we have defined 
\begin{equation} 
\Delta_{n,\ell,m} \equiv \nu_{0,n,\ell}- \nu_{0,n-1,\ell+2}
\end{equation}

Note that in the general case of  $m\not=0$  components, we can define the small separations as
$$d_{n,\ell,m} \equiv {1\over 2}\Bigl(\nu_{n,\ell,m}+\nu_{n,\ell,-m}- \nu_{n-1,\ell+2,m} - \nu_{n-1,\ell+2,-m})\Bigr)$$
which provide  expressions similar to Eq.(5) for each $m$.

For high frequency $p$-modes such as those detected in the Sun and a few other stars,   
 modes  $(n,\ell)$ and $(n-1, \ell+2)$ are systematically 
near degenerate if the star is moderately rotating (Soufi et al 1998) {\sl i.e.} 
their frequency differences are so small that 
a description with a single spherical  harmonics for each mode   is no longer valid
(Chandrasekhar \& Lebovitz, 1962).  Each of the two 
near degenerate modes must be described as a combination of the spherical harmonics  
with degrees $\ell$ and $\ell+2$.
Their frequencies become:
\begin{equation}
\hat \nu_{n,\ell,m}= {1\over 2} \Bigl(\nu_{n,\ell,m}+ \nu_{n-1,\ell+2,m}- 
\sqrt{d^2_{n,\ell,m}+4 H^2_{n,\ell,m}}\Bigr)
\end{equation}
and
\begin{equation}
\hat \nu_{n-1,\ell+2,m}= {1\over 2}\Bigl(\nu_{n,\ell,m}+ \nu_{n-1,\ell+2,m}+ \sqrt{d^2_{n,\ell,m}
+4 H^2_{n,\ell,m}}\Bigr)
\end{equation}
where $H_{n,\ell,m}$ is a coupling term between both near degenerate modes; $\nu_{n,\ell,m}$ is given by Eq.3 and 
 $d_{n,\ell,m}$ by Eq.4.

Hence for a rotating star, the {\it observed } small separation of 
near degenerate $m=0$ modes  is given by: 
\begin{eqnarray}
\hat  d_{n,\ell}&=& \hat \nu_{n,\ell,0}-\hat \nu_{n-1,\ell+2,0}
  \label{eq:equation2}
\end{eqnarray}
where   $\hat d_{n,\ell}$  differs from the small separation
$d^{(0)}_{n,\ell}$    because the oscillation frequencies  are modified by rotation 
according to Eq.(3-7-8).
For later purpose, we note that Eq.7-8-9 imply
\begin{equation}
\hat  d^2_{n,\ell} =d^2_{n,\ell,0} + 4 H^2_{n,\ell,0} 
\end{equation}
These changes due to rotation are quantified below 
for $1.40 M_\odot$ models  with solar chemical composition and no 
overshooting. The stellar  models have been computed,  assuming either no rotation or  
an initial  rotational velocity $v=10$, $15$, $20$, $30$ and $35 $ km/s. 
Ages of the models have been chosen so that they have the same effective temperature ($\log ~T_{\rm  eff}=3.810$) 
hence approximately fall at the same location in the HR diagram (changes in luminosity are negligible).
The selected models are characterized by rotational  
velocities $v= 9.9, 14.8, 19.7, 29.6, 34.6$ km/s , rotation rates 
$\Omega= (0.82, 1.23, 1.64, 2.47, 2.88)$ $10^{-5}$ rad/s 
and ratios $\epsilon= \Omega/(GM/R^3)^{1/2} \sim (2.5, 3.8, 5.1, 8.9)$ $10^{-2}$ respectively.

\begin{figure}[ht]
  \begin{center}
    \epsfig{file=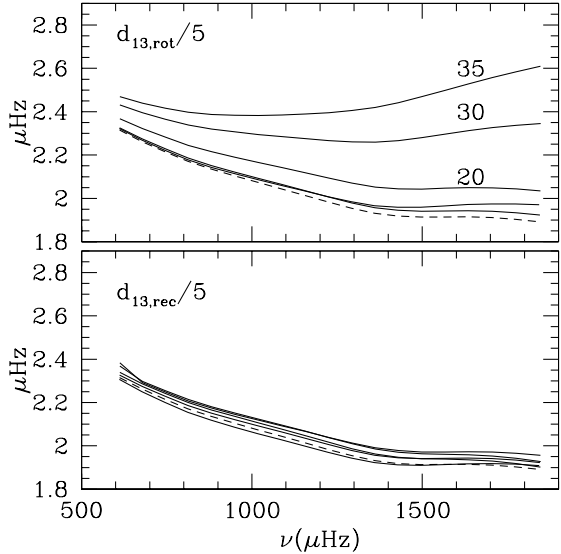, width=\lenA}
  \end{center}
\caption{{\bf top:} Small separations $\hat d_{13}$  
versus frequency  for  $1.4 M_\odot$ models ($\log T_{\rm eff}=3.810$) 
with rotational velocities $v=0$ km/s (dashed line) 
and with $v= 10$, $15$, $20$, $30$, $35$ km/s. The $v\not= 0$ curves depart 
 from the $v=0$ curve by an amount increasing with  $v$ due to rotational effects.
 {\bf bottom:} $d^{(0)}_{13}$ cleaned from  rotational  effects versus frequency. See text for details 
)
 \label{fig4}}
\end{figure}

Oscillation frequencies for these models  have been computed  for modes with degrees $\ell=0-3$ 
in the radial order range $n= 6-25$. This corresponds to a frequency range of about 500-2000 $\mu$Hz.
 Modes $\ell=0,2$ and modes $\ell=1,3$ are considered systematically near 
degenerate and their frequencies are computed according to Eq.7,8.
For each model with initial velocity $v=10$, $15$ ,$20$, $30$ and $35 $ km/s, 
the small separation $\hat d_{13}$   for $m=0$ modes 
is given by  Eq.9  and is shown in  Figure~\ref{fig4}  (top pannel)
 in function of the frequency $\hat \nu_{n,\ell,0}$   for each model.
The effect of rotation - which is mainly due to centrifugal distorsion- can clearly be seen. 
The curves are  seen to deviate from the $v=0$  one, 
the departure  increases with $\Omega^2$ as expected. 
At small enough $v$, only the highest frequency part is affected 
but for higher velocities $v >15$ km/s the whole range of modes is affected. 

As the effect of rotation is to increase the small separation,  
 the point corresponding to the rotating model in the small separation  versus large separation (CD) diagram 
(Christensen-Dalsgaard, 1993)   
would indicate  a star which is younger than it is in reality.   Hence a CD diagram,
 in addition to the chemical composition, must add the rotation as a parameter.

For a $v =20 ~km/s$ for instance,  the small separation is increased by roughly  0.15 $\mu$Hz. 
This departure is of the same order of magnitude as the 
the variation of the small separation with frequency
 which provides information about the 
inner layers and require a precision of $\sim 0.2 \mu$Hz on individual frequencies {\sl i.e.}
which corresponds to  $\sim $ 3 months of observation.
Hence  rotation must be taken into account when 
the star rotates fast ($v >15$ km/s for solar-like oscillating stars).

\subsection { Recovering small separations free of rotational \ effects}

Fortunately, it is  possible to remove the disturbing effects of 
the rotation on the small separation and  to recover the small  separation $d^{(0)}_{n\ell}$ (Eq(2))
free of rotational effects from the observed frequencies (Eq.7-8).

For solar-like oscillations ie with frequencies following  asymptotic relations, 
  a good approximation when the rotation is moderate is 
\begin{equation}
d^{(0)}_{n,\ell}\sim \Delta_{n,\ell,0}
\end{equation} 
This means that  the contribution of the spherically symmetric distorsion of the equilibrium model 
is not removed but is quite negligible here.

As solar-like modes are predominantly acoustic, the coefficients in Eq.(3),(7), and (8) 
can be approximated by
 \begin{equation}
{\cal D}_{n\ell,m} \approx  Q_{\ell,m} ~V_{n,\ell}  ~~~~~~~~ 
 H_{n\ell,m} = Q_{\ell,\ell+2,m} ~V_{n,\ell}
\end{equation}
The coefficient  $V_{n,\ell}$ is roughly the same for both near degenerate modes {\sl i.e.} 
$V_{n,\ell} \sim V_{n-1,\ell+2}$. Geometrical factors are 
 \begin{equation} 
 Q_{\ell,m} = {\Lambda-3 m^2\over 4\Lambda-3 } ~~~~~~~ 
Q_{\ell,\ell+2,m} = {3\over 2} ~\beta_\ell ~\beta_{\ell+2} 
\end{equation}
with 
\begin{equation} 
\Lambda= \ell(\ell+1) ~~~~~~~~~~~~~\beta_\ell= \Bigl({\ell^2- m^2\over 4 \ell^2-1 }\Bigr)^{1/2}   
\end{equation}

From Eq.(5), one derive 
\begin{equation}
\Delta_{n,\ell,0}= d_{n,\ell,0} -\Bigl(Q_{\ell,0}-Q_{\ell+2,0}\Bigr) ~V_{n,\ell} 
\end{equation}
where  the difference  $d_{n,\ell,0}$- using Eq.10-  is given by :
\begin{equation}
 d_{n,\ell,0} = \Bigl(\hat d^2_{n,\ell,0}-4 ~H^2_{n,\ell,0}\Bigr)^{1/2}
\end{equation}
i.e.
\begin{equation}
 d_{n,\ell,0} = \Bigl(\hat d^2_{n,\ell,0}-4 ~ Q^2_{\ell,\ell+2,0} ~V^2_{n,\ell}\Bigr)^{1/2}
\end{equation}

With the help of Eq.(11-15), we can calculate  the small separation  
$d^{(0)}_{n,\ell}$ for two near degenerate modes, $(n,\ell,m=0)$ and  $(n-1,\ell+2,m=0)$,  which is free of
 rotational effects  and  which depends only on
 the observed frequencies and the 
coefficient $V_{n,\ell }$. This coefficient  can in turn be obtained from the observed frequencies. Let define: 
\begin{equation}
s_{n,\ell,m}= {\hat \nu_{n,\ell,m}+ \hat \nu_{n,\ell,-m}\over 2}
\end{equation}

For asymptotic acoustic modes, one then derive for  $\ell=0,2$ near degenerate modes
\begin{equation}
V_{n,0} = {s_{n-1,2,1}-s_{n-1,2,2}\over Q_{2,1}-Q_{2,2}}
\end{equation}
and for $\ell=1,3$  near degenerate modes
\begin{equation}
V_{n,1} = {s_{n-1,3,2}-s_{n-1,3,3}\over Q_{3,2}-Q_{3,3}}
\end{equation}

Figure~\ref{fig4} (bottom pannel) 
 illustrates how efficiently the small separation  $d^{(0)}_{n,\ell}$  (Eq.2) 
can be recovered from the observed frequencies 
$\hat \nu_{n\ell}$ (Eq.(4,5) for a rotating star using Eq.(15), (17) and  (20).
All  $v\not= 0$   curves (recovered) $d^{(0)}_{1,3}/5$  approximately coincide with the  $v=0$ one.

\begin{figure}[ht]
  \begin{center}
    \epsfig{file=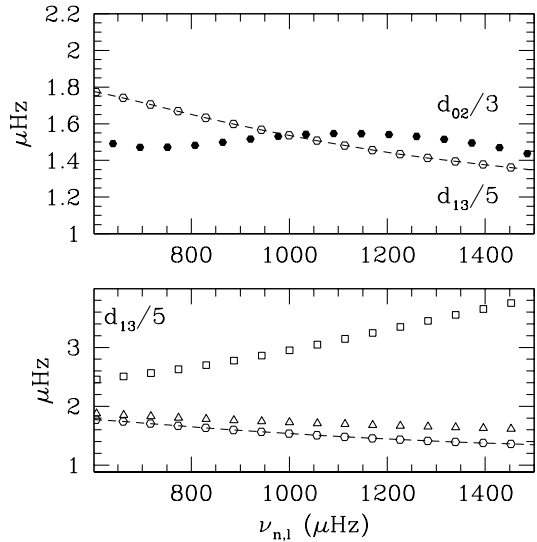, width=\lenA}
  \end{center}
\caption{{\bf top:} Small separations $d_{0,2}/3$ (filled circles) and $d_{1,3}/5$ (dashed line with open circles) 
for a 1.54 $M_\odot$ ($\log T_{\rm eff}=3.820$)
without rotation  versus frequency both in $\mu$Hz
 {\bf Bottom:} $d_{1,3}/5$  with no rotation  as top (dashed line with open circles);  
open  squares : $\hat d_{1,3}/5$ including effects of near-degeneracy   
 for a equivalent model but with a rotational velocity $v= 55.6$ km/s; 
open triangles: $d_{13}/5$ recovered after  removing   contamination by rotational effects.
Note the change of vertical scale between the  top and bottom panels\label{fig5} }
\end{figure}

As a second example, we built a model equivalent to
 Roxburgh's model in the Eddington  assessment report (Favata et al.,  2000) ie a $1.54 M_\odot$ with no overshooting 
and an equivalent evolutionary main sequence  stage.  Figure~\ref{fig5} (top panel)
shows the small separations $d^{(0)}_{0,2}/3$ and $d^{(0)}_{1,3}/5$ for a nonrotating model, 
 which are  quite    comparable with the same quantities for  Roxburgh's model in  Fig.2.7.

 Figure~\ref{fig5} (bottom panel) shows the small separation  $\hat d_{1,3}/5$  
for a $1.54 M_\odot $ model built with  the same effective temperature but 
with a uniform rotation rate $\Omega= 3.94 ~10^{-5}$ rad/s corresponding 
to a rotational velocity   $v=55.6$ km/s (and $\epsilon= 0.15$).  
The small separation calculated with the observed frequencies 
according to Eq. (7-9) significantly differs from the small separation in absence of rotation. 
The difference increases with the frequency and reaches 2.2 $\mu$Hz at $\sim 1500 \mu$Hz.  However, 
combining the frequencies as described above ie using Eq.(11), (15) and (17), it is found that one can remove 
most effects of rotation   and recover a small separation free of these contaminations at the level $< 0.2 \mu$Hz
as can be seen in Figure~\ref{fig5}.

\subsection {Ideal case for inversion of rotation profile}

For stars with opacity driven modes, 
it is expected that one can  obtain detailed information about the rotation profile inside the 
stars providing a few appropriate modes are excited, detected and identified  (Goupil et al., 1996).

For stars with stochastically excited modes, it is known that 
the asymptotic p-modes can hardly give very localized information  about rotation  
everywhere inside the star (Gough, 1998, Goupil el al., 1996). 
However, this is true only if the excited modes are of purely acoustic nature. 
In more evolved stars, among the modes 
that may be stochastically excited, some may have a 
mixed character and thus exhibit some information about the structure and rotation in the inner layers.

\begin{figure}[ht]
  \begin{center}
    \epsfig{file=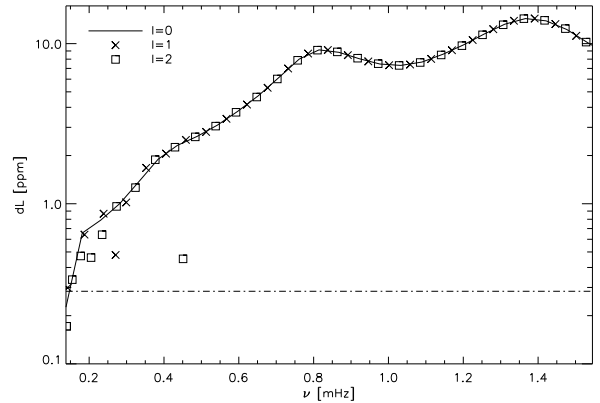, width=\lenA}
  \end{center}
\caption{Expected relative luminosity fluctuations $\delta L/L$ for a $M=1.60~M_\odot$ main sequence stellar model. 
The solid line corresponds to the amplitude of the radial modes ($\ell=0$), 
the crosses to the $\ell=1$ modes and the squares to the $\ell=2$ modes. 
The horizontal dot dashed line corresponds to the  detection threshold expected with  
the Eddington mission  for an observing period of 30 days and a star with magnitude $m_v=6$.
Points departing downward from the $\ell=0$ line corespond to the mixed modes. Note that a few of 
them lie above the Eddington 
threshold}
\label{excited mixed modes}
\end{figure}

Fig. \ref{excited mixed modes}  shows the expected distribution of 
the oscillation amplitudes  (luminosity amplitude $\delta L/L$) 
of stochastically excited modes  with frequency  for a 1.60 $M_\odot$  main sequence model with a  
central hydrogen  abundance $X_c=0.2$.

The stellar model is built with B\"ohm-Vitense (1958)'s formulation 
of the mixing-length (MLT) and assumes the Eddington classical gray atmosphere.
The eigenmodes and eigenfrequencies are calculated with the 
adiabatic ADIPLS pulsation code (Christensen-Dalsgaard, 1996). 
The calculations of the mode amplitudes make use of  the adiabatic assumption for relating $\delta L/L$ 
to the rates $P$ and the rates $\eta$ at which the mode are excited and damped respectively. 
The excitation rates $P$ are computed according to the stochastic excitation model of Samadi \& Goupil (2001). 
The calculations assume -~as in Samadi et al. (2003) ~- a Lorenzian function for modelling 
the convective eddy  time-correlations.
Using the same physics  (MLT)  for a  solar model leads to an underestimation of 
the excitation power $P$ compared with the observations. In order  for  the

\noindent  maximum in $P$ to match that of  
Chaplin et al. (1998)'s seismic observations, 
the solar excitation rates $P$ must be   multiplied by a factor 13. Accordingly, we multiply by the same factor the 
computed excitation rates for the stellar model considered here. 
The damping  rates $\eta$ are obtained from Houdek et al (1999) 's tables.

 Figure~\ref{excited mixed modes} displays the amplitudes of  
luminosity fluctuations $\delta L/L$ for each mode  
in the frequency range expected to be sensitive  to a  stochastic excitation by turbulent convection 
for a $M=1.60~M_\odot$ main sequence star. 
The  Eddington detection threshold is also plotted. From the comparison, one expects 
many modes to be detected for such a star. Among these detected modes, a few are mixed modes which 
are seen to deviate from the general trend 
in Figure~\ref{excited mixed modes}. These modes have amplitudes
 smaller than the other modes in the same frequency interval; 
this is due to their mixed nature, they have amplitudes in the inner layers 
which increase their inertia compared to that of the  neighboring pure $p$-modes.  

Assuming that rotational splitting of all mixed modes with amplitudes above the threshold is measured, 
 Lochard et al. (2003 in prep) found that rotation rates in the $\mu$-gradient zone and in the chemically 
homogeneous envelope could be determined with some precision. 
This would provide a valuable constraint on theories of angular momentum transport
in stellar interiors

\end{document}